\documentclass[12pt]{article}
\usepackage{bm}
\begin{document}
\title{Left-Right Symmetry Model with Two Bidoublets and One Doublet Higgs Fields for Electroweak Interaction}
\date{}
\maketitle
\begin{center}
\textbf{Asan Damanik\footnote{E-mail: d.asan@lycos.com}}\\
Department of Physics, Sanata Dharma University,\\ Kampus III USD Paingan, Maguwoharjo, Sleman,Yogyakarta, Indonesia\\
and Graduate School, Gadjah Mada University, Yogyakarta,Indonesia\\
\end{center}
\begin{center}
\textbf{M. Satriawan} and \textbf{Muslim\footnote{Deceased on May 27,2007}}\\
Department of Physics, Gadjah Mada University, Yogyakarta, Indonesia\\
\end{center}
\begin{center}
\textbf{Pramudita Anggraita}\\
National Nuclear Energy Agency (BATAN), Jakarta, Indonesia
\end{center}     
\abstract{We use the left-right symmetry model based on $SU(2)_{L}\otimes SU(2)_{R}\otimes U(1)_{B-L}$ gauge group with two bidoublets and one doublet Higgs fields for electroweak interaction.  The lepton fields are represented as a doublet of SU(2) for both left and right fields.   By using the pattern of the symmetry breaking emerges as the minimum of the Higgs potential for a range of parameters, we show that the domination of the V-A over V+A interactions is natural at the low energy. The symmetry breaking is the responsible mechanism for the \textit{up-down} lepton doublet mass difference.}    

\section{Introduction}
Even though the Glashow-Weinberg-Salam (GWS) model for electroweak interaction (standard model for electroweak interaction) which is based on $SU(2)_{L}\otimes U(1)_{Y}$ gauge symmetry group (see for example Peskin and Schroeder \cite{Peskin95}) has been in succes phenomenologically, but it still far from a complete theory because the theory does not explain many fundamental problems such as neutrino mass problem, and fermions (lepton and quark) mass hierarchy \cite{Fukugita03}.  Recent experimental data on atmosphere and solar neutrinos indicate strongly that the neutrinos are massive and mixed up one another \cite{Fukuda98, Fukuda99, Fukuda01,Toshito01, Giacomelli01,Ahmad02, Ahn03}. Many theories or models have been proposed to extend the standard model of electroweak interaction.  One of the interesting model is the left-right symmetry model based on the $SU(2)_{L}\otimes SU(2)_{R}\otimes U(1)$ gauge group which is proposed by Senjanovic and Mohapatra \cite{Senjanovic75}.

By intoducing two doublets Higgs fields, Senjanovic and Mohapatra found that the Higgs potential has to be minimum when we choose the asymmetric solution to the doublet Higgs fields vacuum expectation values.  Within this scheme, the presence of the spontaneous parity violation at low energy is natural and the electroweak interaction based on $SU(2)_{L}\otimes U(1)_{Y}$ gauge symmetry group can be deduced from the left-right symmetry model and then it breaks into $U(1)_{em}$ just like GWS model.  By introducing one additional bidoublet Higgs fields to break $SU(2)_{L}\otimes SU(2)_{R}\otimes U(1)$ down to $U(1)_{em}$, it was shown that there are two massive gauge bosons $m_{W_{L}}$ and $m_{W_{R}}$ where $m_{W_{L}}$ $<<$ $m_{W_{R}}$, and a small neutrino mass for left-handed neutrinoes could be produced via seesaw mechanism.

Siringo \cite{Siringo04} noticed that any viable gauge model for electroweak interactions must give an answer to the two quite different problems: (i) the breaking of symmetry from the full gauge group into electromagnetic Abelian group $U(1)_{em}$ giving a mass to the gauge bosons and then explains the known structure of weak interactions, and (ii) the mass matrices for fermions.  In his other paper \cite{Siringo03}, by imposing the $O(2)$ custodial symmetry with left and right Higgs fields chosen to be a doublet of $SU(2)$, Siringo obtained the known \textit{up-down} structure of the doublet fermions masses by insertion of \textit{ad hoc} fermion-Higgs interactions.  Meanwhile, Montero and Pleitez \cite{Montero06} used the approximate custodial $SU(2)_{L+R}$  global symmetry to extend the GWS model. But, the $O(2)$ custodial symmetry leads to five dimensions operator in the mass term of the Lagrangian density which is not renormalizable.   A theory is renormalizable if the dimension of the operator in the Lagrangian density less than or equal to 4 \cite{Mandl84, Ryder85}.  Thus, existency of the $O(2)$ custodial symmetry in the the electroweak theory is still a problem.

In this paper, we use the left-right symmetry model based on $SU(2)_{L}\otimes SU(2)_{R}\otimes U(1)_{B-L}$ gauge group by using two bidoublets and one doublet Higgs fields to break the left-right symmetry based on $SU(2)_{L}\otimes SU(R)\otimes U(1)_{B-L}$ gauge group down to $U(1)_{em}$ and then we study the predictive power of the model to the fundamental procesess.  The fermion fields are chosen to be an $SU(2)$ doublet for both left and right fields.  Thus, in Section 2 we introduce explicitly our model and evaluate the minimum Higgs field potential by choosing the appropriate vacuum expectation values of the Higgs fileds.  In section 3, we explicitly give a systematic calculations of the gauge bosons and fermions masses.  In section 4, we discuss our results and its implications to the fundamental processes, and finally in section 5 we present a conclusion.

\section{The Model}
The left-right symmetry model based on $SU(2)_{L}\otimes SU(2)_{R}\otimes U(1)_{B-L}$ with the following lepton fields assignment:
\begin{eqnarray}
\psi_{L}=\bordermatrix{&\cr
&\nu_{l}\cr
&l^-\cr}_{L}, \  \psi_{R}=\bordermatrix{&\cr
&\nu_{l}\cr
&l^-\cr}_{R}
 \label{eq:lepton}
\end{eqnarray}
where $l=e,\mu,\tau$, and three Higgs fields (two bidoublets and one doublet fields) with its electromagnetic charges reads:
\begin{eqnarray}
\phi_{1}=\bordermatrix{& &\cr
&a^0 &b^+\cr
&c^- &d^0\cr},\ \phi_{2}=\bordermatrix{& &\cr
&e^0 &f^+\cr
&g^- &h^0\cr},\nonumber\\ \Phi=\bordermatrix{&\cr
&p^+\cr
&q^0\cr}\ \ \ \ \ \ \ \ \ \ \ \ \ \ \ \ 
  \label{eq:higgs}
\end{eqnarray}
which break the $SU(2)_{L}\otimes SU(2)_{R}\otimes U(1)_{B-L}$ down to $U(1)_{em}$.  The bidoublet Higgs $\phi_{1}$ transforms as $\phi_{1}(2,2,-2)$, $\phi_{2}$ transforms as $\phi_{2}(2,2,-2)$, and $\Phi$ transforma as $\Phi(0,2,0)$ under $SU(2)$ respectively.  The general potential of the Higgs fields which consistent with renormalizability, gauge invariance, and discrete left-right symmetry is the following:
\begin{eqnarray}
V(\phi_{1},\phi_{2})=-\mu^2\left[Tr(\phi_{1}^+\phi_{1})+Tr(\phi_{2}^+\phi_{2})\right]\nonumber\\+\lambda_{1}[(Tr(\phi_{1}^+\phi_{1})^2+(Tr(\phi_{2}^+\phi_{2})^2]\nonumber\\+\lambda_{2}[Tr(\phi_{1}^+\phi_{1})Tr(\phi_{2}^+\phi_{2})]\nonumber\\-\alpha^2(\Phi^+\Phi)+\beta(\Phi^+\Phi)^2\nonumber\\+\gamma\left[Tr(\phi_{1}^+\phi_{1})+Tr(\phi_{2}^+\phi_{2})\right](\Phi^+\Phi)\nonumber\\+\delta\left[Tr(\phi_{1}^+\Phi\Phi^+\phi_{1})+Tr(\phi_{2}^+\Phi\Phi^+\phi_{2})\right]+H.c.
   \label{eq:potential}
\end{eqnarray}

After explicitly performing the calculation to find out the minimum potential of the Higgs fields in Eq.(\ref{eq:potential}), we obtain the following constraints:
\begin{eqnarray}
Tr(\phi_{1}^+\phi_{1})=Tr(\phi_{2}^+\phi_{2})=\frac{2\mu^2-2(\gamma+\delta)\Phi^+\Phi}{2\lambda_{1}+\lambda_{2}},
  \label{eq:const1}
\end{eqnarray}
and
\begin{eqnarray}
\Phi^+\Phi=\frac{\left[\alpha^2-2(\beta+\gamma)\right]Tr(\phi_{1}^+\phi_{1})}{2\beta}.
  \label{eq:const2}
\end{eqnarray}
From Eqs.(\ref{eq:const1}) and (\ref{eq:const2}), we can see that the minimum potential $V(\phi_{1},\phi_{2},\Phi)$ in Eq.(\ref{eq:potential}) can be made to be minimum by arranging the values of the parameters $\lambda_{1}, \lambda_{2}, \beta, \gamma$, and $\delta$.

We can choose the minimum so that it has vacuum expectation values of the Higgs fields, for the domain of parameters $\mu^2>\alpha^2,  \lambda_{2}>2\lambda_{1}$, and $e<<h$ as follow:
\begin{eqnarray}
\left\langle\phi_{1}\right\rangle=\bordermatrix{& &\cr
&0 &0\cr
&0 &0\cr},\ \left\langle\phi_{2}\right\rangle=\bordermatrix{& &\cr
&e &0\cr
&0 &h\cr},\nonumber\\ \left\langle\Phi\right\rangle=\bordermatrix{&\cr
&0\cr
&q\cr}\ \ \ \ \ \ \ \ \ \ \ \ \ 
  \label{eq:vev}
\end{eqnarray}
This set is known as an asymmetric solution. The asymmetric solution guarantee the presence of parity violation at low energy as known today in the electroweak interactions phenomenologically.

The complete Lagrangian density $L$ in our model is given by:
\begin{eqnarray}
L=-\frac{1}{4}W_{\mu\nu L}.W^{\mu\nu L}-\frac{1}{4}W_{\mu\nu R}.W^{\mu\nu R}-\frac{1}{4}B_{\mu\nu}B^{\mu\nu}\nonumber\\+\bar{\psi_{L}}\gamma^{\mu}\left(i\partial_{\mu}-g\frac{1}{2}\tau.W_{\mu L}-g'\frac{B-L}{2}B_{\mu}\right)\psi_{L}\nonumber\\+\bar{\psi_{R}}\gamma^{\mu}\left(i\partial_{\mu}-g\frac{1}{2}\tau.W_{\mu R}-g'\frac{B-L}{2}B_{\mu}\right)\psi_{R}\nonumber\\+Tr\left|\left(i\partial_{\mu}-g\frac{1}{2}\tau.W_{\mu L}-g'\frac{B-L}{2}B_{\mu}\right)\phi_{1}\right|^2\nonumber\\+Tr\left|\left(i\partial_{\mu}-g\frac{1}{2}\tau.W_{\mu R}-g'\frac{B-L}{2}B_{\mu}\right)\phi_{1}\right|^2\nonumber\\+Tr\left|\left(i\partial_{\mu}-g\frac{1}{2}\tau.W_{\mu L}-g'\frac{B-L}{2}B_{\mu}\right)\phi_{2}\right|^2\nonumber\\+Tr\left|\left(i\partial_{\mu}-g\frac{1}{2}\tau.W_{\mu R}-g'\frac{B-L}{2}B_{\mu}\right)\phi_{2}\right|^2\nonumber\\+\left|\left(i\partial_{\mu}-g\frac{1}{2}\tau.W_{\mu R}-g'\frac{B-L}{2}B_{\mu}\right)\Phi\right|^2\nonumber\\-V(\phi_{1},\phi_{2},\Phi)-(m_{o}\bar{\psi_{L}}\psi_{R}+G_{1l}\bar{\psi_{L}}\phi_{1}\psi_{R}\nonumber\\+G_{2l}\bar{\psi_{L}}\phi_{2}\psi_{R}+H.c.)\ \ \ \ \ \ \ \ \ \ \ \ \ \ \ \ \ \ \ \ \ \ \ \ \ 
  \label{eq:lagrangian}
\end{eqnarray}
where $\gamma^{\nu}$ is the usual Dirac matrices, $\tau$ is the Pauli spin matrices, $g$ is the $SU(2)$ coupling , $g'$ is the $U(1)$ coupling, $m_{o}$ is the free lepton mass, $G_{1l}$ and $G_{2l}$ are the Yukawa couplings, and the $B-L$ quantum number is associated with the $U(1)$ generator in the left-right symmetry model.  The relation of the $B-L$ to the electric charge $Q$ in the left-right symmetry model is given by \cite{Mohapatra03}:
\begin{eqnarray}
Q=T_{3L}+T_{3R}+\frac{B-L}{2}
  \label{eq:charge}
\end{eqnarray}

\section{The Gauge Bosons and Leptons Masses}
\subsection{Gauge Bosons Masses}
The relevant mass terms for gauge bosons can be obtained from Eq.(\ref{eq:lagrangian}), they are the sixth, seventh, eighth, ninth, and the tenth terms without $i\partial_{\mu}$. By substituting Eq.(\ref{eq:vev}) into these relevant mass terms, we obtain:
\begin{eqnarray}
Tr\left|\left(-g\frac{1}{2}\tau.W_{\mu L}-\frac{g'}{2}B_{\mu}\right)\left\langle \phi_{1}\right\rangle\right|^2+Tr\left|\left(-g\frac{1}{2}\tau.W_{\mu R}-\frac{g'}{2}B_{\mu}\right)\left\langle \phi_{1}\right\rangle\right|^2\nonumber\\+Tr\left|\left(-g\frac{1}{2}\tau.W_{\mu L}-\frac{g'}{2}B_{\mu}\right)\left\langle \phi_{2}\right\rangle\right|^2+Tr\left|\left(-g\frac{1}{2}\tau.W_{\mu R}-\frac{g'}{2}B_{\mu}\right)\left\langle \phi_{2}\right\rangle\right|^2\nonumber\\+\left|\left(-g\frac{1}{2}\tau.W_{\mu R}-\frac{g'}{2}B_{\mu}\right)\left\langle \Phi\right\rangle\right|^2\nonumber\\=\frac{A^2g^2}{4}\left\{(W_{\mu R}^1)^2+(W_{\mu R}^2)^2\right\}+\frac{D^2g^2}{4}\left\{(W_{\mu L}^1)^2+(W_{\mu L}^2)^2\right\}\nonumber\\+\frac{e^2}{4}\left(gW_{\mu L}^3-2g'B_{\mu}\right)^2+\frac{h^2}{4}\left(gW_{\mu L}^3+2g'B_{\mu}\right)^2\nonumber\\+\frac{e^2}{4}\left(gW_{\mu R}^3-2g'B_{\mu}\right)^2+\frac{h^2}{4}\left(gW_{\mu R}^3+2g'B_{\mu}\right)^2+\frac{q^2g^2}{4}\left(W_{\mu R}^3\right)^2,
  \label{eq:bosonmass}
\end{eqnarray}
where $A^2=e^2+h^2+q^2,\;D^2=e^2+h^2$, and we have taken the value of $B-L=-2$ for both $\phi_{1}$ and $\phi_{2}$ and $B-L=0$ for $\Phi$ to satisfy the requirement that the vacuum is invariant under $U(1)_{em}$ transformation and the photon remain massless.

To find out the gauge bosons masses explicitly, we define:
\begin{eqnarray}
W_{\mu j}^\pm=\frac{1}{\sqrt2}(W_{\mu j}^1 \mp iW_{\mu j}^2),\ \ \ \ \ \ \ \ \ \ \ \ \ \ \ \ \ \ \nonumber\\Z_{1\mu j}=\frac{gW_{\mu j}^3-2g'B_{\mu}}{\sqrt{g^2+4g'^2}}=W_{\mu j}^3 \cos{\theta_{W}}-B_{\mu}\sin{\theta_{W}},\nonumber\\Z_{2\mu j}=\frac{gW_{\mu j}^3+2g'B_{\mu}}{\sqrt{g^2+4g'^2}}=W_{\mu j}^3 \cos \theta_{W}+B_{\mu} \sin \theta_{W},\nonumber\\A_{1\mu j}=\frac{2g'W_{\mu j}^3+gB_{\mu}}{\sqrt{g^2+4g'^2}}=W_{\mu j}^3 \sin{\theta_{W}}+B_{\mu}\cos{\theta_{W}},\nonumber\\A_{2\mu j}=\frac{2g'W_{\mu j}^3-gB_{\mu}}{\sqrt{g^2+4g'^2}}=W_{\mu j}^3 \sin \theta_{W}-B_{\mu} \cos \theta_{W},\nonumber\\X_{\mu R}=W_{\mu R}^3,\ \ \ \ \ \ \ \ \ \ \ \ \ \ \ \ \ \ \ \ \ \ \ \ \ 
  \label{eq:parameter}
\end{eqnarray}
where $j=L,R$, and we have introduced $\theta_{W}$ (the weak mixing angle) defined by:
\begin{eqnarray}
\cos\theta_{W}=\frac{g}{\sqrt{g^2+4g'^2}},\ \  \sin\theta_{W}=\frac{2g'}{\sqrt{g^2+4g'^2}}.
  \label{eq:mixingangle}
\end{eqnarray}
By substituting Eqs.(\ref{eq:parameter}) and (\ref{eq:mixingangle}) into Eq.(\ref{eq:bosonmass}), we finally obtain the relevant mass terms in the Lagrangian density of the physical $W_{\mu R}^\pm, \;W_{\mu L}^\pm, \;Z_{1\mu R}, \;Z_{2\mu R}, \;Z_{1\mu L},\;Z_{2\mu L}$, $\;X_{\mu R}$, and $A_{\mu}$ fields, namely:
\begin{eqnarray}
\frac{1}{2}m_{W_{R}}^2W_{\mu R}^+W_{R}^{-\mu}+\frac{1}{2}m_{Z_{1R}}^2Z_{1\mu R}^2+\frac{1}{2}m_{Z_{2R}}^2Z_{2\mu R}^2+\frac{1}{2}m_{x}^2X_{\mu }^2\nonumber\\+\frac{1}{2}m_{\gamma}^2A_{\mu }^2+\frac{1}{2}m_{W_{L}}^2W_{\mu L}^+W_{L}^{-\mu}+\frac{1}{2}m_{Z_{1L}}^2Z_{1\mu L}^2+\frac{1}{2}m_{Z_{2L}}^2Z_{2\mu L}^2.
\end{eqnarray}
where:
\begin{eqnarray}
m_{W_{R}}=\frac{1}{\sqrt{2}}Ag,\ \  m_{Z_{1R}}=\frac{e}{\sqrt{2}}\left(g^2+4g'^2\right)^{\frac{1}{2}},m_{Z_{2R}}=\frac{h}{\sqrt{2}}\left(g^2+4g'^2\right)^{\frac{1}{2}},\nonumber\\m_{W_{L}}=\frac{Dg}{\sqrt{2}},\ \  m_{Z_{1L}}=\frac{e}{\sqrt{2}}\left(g^2+4g'^2\right)^{\frac{1}{2}},m_{Z_{2L}}=\frac{h}{\sqrt{2}}\left(g^2+4g'^2\right)^{\frac{1}{2}}, \nonumber\\m_{x}=\frac{qg}{\sqrt{2}},\ \ m_{\gamma}=0.\ \ \ \ \ \ \ \ \ \ \ \ \ \ \ \ \ \ \ \ \ \ \ 
  \label{eq:bosonmass1}
\end{eqnarray}

From Eq.(\ref{eq:bosonmass1}) we see that the photon is massless ($m_{\gamma}=0$), and the bosons $W, \;X$, and $Z$ are massive with $m_{W_{R}} > m_{W_{L}}$ and $m_{Z_{R}} = m_{Z_{L}}$.
\subsection{Leptons Masses}
As one can read from Eq.(\ref{eq:lagrangian}), the leptons masses term in the Lagrangian density is: 
\begin{eqnarray}
L_{mass}=m_{o}\bar{\psi_{L}}\psi_{R}+G_{1l}\bar{\psi_{L}}\phi_{1}\psi_{R}\nonumber\\+G_{2l}\bar{\psi_{L}}\phi_{2}\psi_{R}+H.c.
 \label{eq:leptonmass}
\end{eqnarray}
From Eq.(\ref{eq:leptonmass}),one can see that the leptons mass terms in our model arise from two kinds of mass term, they are: (i) free lepton mass term, and (ii) Yukawa term.  The free lepton mass term is the mass of the lepton as a free particle, and the Yukawa term is an additional mass to the free lepton mass which is resulted from symmetry breaking of the gauge fields.  

By substituting Eqs.(\ref{eq:lepton}) and (\ref{eq:vev}) into Eq.(\ref{eq:leptonmass}), we obtain:
\begin{eqnarray}
L_{mass}=m_{o}\left[\bar{\nu}_{lL}\nu_{lR}+\bar{l}_{L}{l}_{R}\right]\ \ \ \ \ \ \ \ \ \ \ \ \ \  \nonumber\\+G_{2l}\left[e\bar{\nu}_{lL}\nu_{lR}+h\bar{l}_{L}{l}_{R}\right]+H.c.\nonumber\\=\left[m_{o}+G_{2l}e\right]\bar{\nu}_{lL}\nu_{lR}\ \ \ \ \ \ \ \ \ \ \ \ \ \ \  \nonumber\\+\left[m_{o}+G_{2l}h\right]\bar{l}_{L}{l}_{R}+H.c.\ \ \ \ \ \ \ 
  \label{eq:leptonmass1}
\end{eqnarray}
where $m_{o}+G_{2l}e=m_{\nu_{l}}$ is the neutrino mass, and $m_{o}+G_{2l}h=m_{l}$ is the electron or muon or tauon mass.  Since the value of the $e << h$, it implies that the values  of the $m_{\nu_{l}}<< m_{l}$ as known today.
\section{Discussion}
By using the left-right symmetry model based on $SU(2)_{L}\otimes SU(2)_{R}\otimes U(1)_{B-L}$ gauge group with two bidoublets and one doublet Higgs fields, we obtain eight bosons (seven bosons to be massive and one massless) after symmetry breaking.  Two bosons, $W_{R}$ charge boson with mass $m_{W_{R}}$ and $X$ neutral boson with mass $m_{x}$, are very massive.  To see qualitatively the contribution of the $W_{R}$ to the weak interaction at low energy, we can check it via Fermi coupling $G_{F}$. The effective interactions at low energy is proportional to Fermi coupling $G_{F}$.  The relation of the $G_{F}$ to the $W$ boson mass is given by:  
\begin{eqnarray}
G_{F}=\frac{\sqrt{2}g^2}{8m_{W}^2}
\end{eqnarray}
Because the $W_{R}$ boson is very massive compared to the $W_{L}$ boson, then the contribution of the $W_{R}$ boson (with mass $m_{W_{R}}$) interactions with leptons fields at low energy is very weak compared to the interactions of the $W_{L}$ boson (with mass $m_{W_{L}}$) with the leptons fields.  
Thus, the structure of the electroweak interactions as known today which is dominated by the V-A interactions due to the very massiveness of the $W_{R}$ such that the contribution of the V+A interactions are very small.  The $X_{R}$ neutral boson with mass $m_{X_{R}}$ is a new boson.  

From Eqs.(\ref{eq:mixingangle}), and (\ref{eq:bosonmass1}) we obtain a relation:
\begin{eqnarray}
\frac{m_{W_{R}}}{m_{Z_{1R}}}=\frac{A}{e}\cos \theta_{W},\ \ \ \frac{m_{W_{R}}}{m_{Z_{2R}}}=\frac{A}{h}\cos \theta_{W},\nonumber\\ \frac{m_{W_{R}}}{m_{x}}=\frac{A}{q	},\ \ \  \frac{m_{W_{L}}}{m_{Z_{1L}}}=\frac{D}{e}\cos \theta_{W},\ \ \ \frac{m_{W_{L}}}{m_{Z_{2L}}}=\frac{D}{h}\cos \theta_{W}.
\end{eqnarray}
and
\begin{eqnarray}
\tan{\theta_{W}}=\frac{2g'}{g}.
\end{eqnarray}

By using the left-right symmetry model, the neutrino masses arise naturally.  The free lepton masses $m_{0}$ should be zero if we fully adopt the GWS model which dictate that all of the leptons acquire mass after the symmetry breaking.  The problem of the \textit{up-down} doublet mass difference in the lepton sector can be understood qualitatively as one can seen in Eq.(15).  The \textit{up-down} doublet lepton mass difference arises from the additional fermion-Higgs coupling constant $G_{2l}$ with vacuum expectation values of the $\phi_{2}$.  Certainly, this additional term is assigned to the the symmetry breaking.  If the symmetry breaking do not take place, then one can see that the \textit{up-down} mass in the lepton doublet is equal.  Thus, the responsible mechanism  for the difference of the $up-down$ lepton doublet mass is the symmetry breaking. On the other hand, if there is no symmetry breaking, then the \textit{up-down} lepton doublet mass is equal.  The fermion-Higgs coupling $G_{2l}$ is arbitrary just like the fermion-Higgs coupling in the GWS model.
\section{Conclusion}
In the scheme of a left-right symmetry model based on $SU(2)_{L}\otimes SU(2)_{R}\otimes U(1)_{B-L}$ gauge group with two bidoublets and one doublet Higgs fields, and the lepton fields are represented as an $SU(2)$ doublet for both left and right fields,  we obtain two very massive bosons after the symmetry breaking take place, the $W_{R}$ charge boson with mass $m_{W_{R}}$ and the new $X_{R}$ neutral boson with mass $m_{x}$.  Because the $W_{R}$ bosons is very massive compared to the $W_{L}$ boson, thus its interactions with lepton fields are very weak compared to the interactions of the $W_{L}$ boson (with mass $m_{W_{L}}$) with leptons fields.  Thus, the structure of the electroweak interactions which is dominated by the V-A interactions could be understood as due to the very massiveness of the $W_{R}$ boson such that the contribution of the V+A interactions are very small compared to the V-A interaction at low energy. The responsible mechanism  for the $up-down$ lepton doublet mass difference is the symmetry breaking. The neutrino mass arises naturally in the left-right symmetry model.  We also obtained that the weak mixing angle $\theta_{W}=\arctan{\left(\frac{2g'}{g}\right)}$.  
\section*{Acknowledgments}
The first author would like to thank to the Graduate School of Gadjah Mada University Yogyakarta where he is currently a graduate doctoral student, the Dikti Depdiknas for a BPPS Scholarship Program, and the Sanata Dharma University Yogyakarta for granting the study leave and opportunity.


\begin{thebibliography}{99}
\bibitem{Peskin95}
M. E. Peskin and D. V. Schroeder, \textit{An Introduction to Quantum Field Theory}, Addison-Wesley Publishing Company, 1995.
\bibitem{Fukugita03}
M. Fukugita and T. Yanagida, \textit{Physics of Neutrinos and Application to Astrophysics}, Springer-Verlag, Heidelberg, 2003.
\bibitem{Fukuda98}
Y. Fukuda \textit{et al., Phys. Rev. Lett.} {\bf81}, 1158 (1998).
\bibitem{Fukuda99}
Y. Fukuda \textit{et al., Phys. Rev. Lett.} {\bf82}, 2430 (1999).
\bibitem{Fukuda01}
Y. Fukuda \textit{et al., Phys. Rev. Lett.} {\bf 86}, 5656 (2001). 
\bibitem{Toshito01}
T. Toshito \textit{et al., hep-ex}/0105023.
\bibitem{Giacomelli01}
G. Giacomelli and M. Giorgini, \textit{hep-ex}/0110021.
\bibitem{Ahmad02}
Q.R. Ahmad \textit{et al., Phys. Rev. Lett.} {\bf89}, 011301 (2002).
\bibitem{Ahn03}
M. H. Ahn \textit{et al., Phys. Rev. Lett.} {\bf90}, 041801-1 (2003).
\bibitem{Senjanovic75} 
G. Senjanovic and R.N. Mohapatra, \textit{Phys. Rev}.{\bf D12(5)},1502 (1975).
\bibitem{Siringo04}
F. Siringo, \textit{Phys. Rev. Lett}. {\bf 92(11)},119101-1 (2004). 
\bibitem{Siringo03}
F. Siringo, \textit{hep-ph}/0307320 v1.
\bibitem{Montero06}
J.C. Montero and V. Pleitez, \textit{hep-ph}/0607144 v1.
\bibitem{Mandl84}
F. Mandl and G. Shaw, \textit{Quantum Field Theory}, John Wiley {\&} Sons, New York, 1984.
\bibitem{Ryder85}
L. H. Ryder, \textit{Quantum Field Theory}, Cambridge University Press, Cambridge, 1985.
\bibitem{Mohapatra03}
R.N. Mohapatra, \textit{Unification and Supersymmetry: The Frontiers of Quark-Lepton Physics}, Third Edition, Springer, New York,  2003. 
\end{thebibliography}
\end{document}